*Article*

# Effective Medium Theory for the Elastic Properties of Composite Materials with Various Percolation Thresholds

**Andrei A. Snarskii[1, 2, *], Mikhail Shamonin[3,*] and Pavel Yuskevich[1]**

[1] National Technical University of Ukraine "Igor Sikorsky Kyiv Polytechnic Institute", Prospekt Peremohy 37, 03056 Kiev, Ukraine; asnarskii@gmail.com (A.S); pasha.yusk@gmail.com (P.Y.)
[2] Institute for Information Recording, NAS of Ukraine, Mykoly Shpaka Street 2, 03113 Kiev, Ukraine
[3] East Bavarian Centre for Intelligent Materials (EBACIM), Ostbayerische Technische Hochschule (OTH) Regensburg, Seybothstr. 2, D-93053 Regensburg, Germany; mikhail.chamonine@oth-regensburg.de (M.S.)

* Correspondence: asnarskii@gmail.com (A.S); mikhail.chamonine@oth-regensburg.de (M.S.)

**Abstract:** It is discussed that the classical effective medium theory for the elastic properties of random heterogeneous materials is not congruous with the effective medium theory for the electrical conductivity. In particular, when describing the elastic and electro-conductive properties of a strongly inhomogeneous two-phase composite material, the steep rise of effective parameters occurs at different concentrations. To achieve the logical concordance between the cross-property relations, a modification of the effective medium theory of the elastic properties is introduced. It is shown that the qualitative conclusions of the theory do not change, while a possibility of describing a broader class of composite materials with various percolation thresholds arises. It is determined under what conditions there is an elasticity-theory analogue of the Dykhne formula for the effective conductivity. The theoretical results are supported by known experiments and show improvement over the existing approach. The introduction of the theory with the variable percolation threshold paves the way for describing the magneto-elastic properties of magnetorheological elastomers, as it has been achieved for magneto-dielectric and magnetic properties recently.

**Keywords:** elastic properties; effective medium approximation; self-consistent; random heterogeneous medium; two-phase composite material; percolation threshold

---

## 1. Introduction

The calculation of effective physical properties of composite materials is of significant interest for many branches of science and engineering, because it allows one to predict the characteristics of the resulting material from those of its constitutive components. The pioneering papers written by J. C. Maxwell and J. W. S. Rayleigh almost 150 years ago provided correct statement of the problem and a solution for small concentrations of inclusions. Significant progress in understanding the behavior of composite materials has been achieved in the second half of the past century. This progress is associated with the percolation theory, theoretical description of nanostructures and nonlinear properties as well as the development of smart composite materials, see, e.g. [1-4] for the reviews of the subject. The present paper concerns the calculation of effective mechanical properties of random heterogeneous two-phase composite materials, the effective medium theory (EMT) of those was formulated about 55 years ago [5,6]. However, this approximate method is used until now, see e.g. [7-10]. We demonstrate below that that in some cases this classical theory does not behave adequately. For example, it leads to an antilogy in the description of a simultaneously elastic and electro-conductive composite material. For a two-dimensional random heterogeneous medium, the classical EMT yields a non-symmetric expression with respect to the interchange of the mechanical properties of phases. We propose a modification of the traditional EMT for the elastic properties, which allows one to eliminate some inconsistency observable when comparing this EMT for elastic properties with the EMT for other physical properties of the same material. The traditional EMT for the elastic properties of random heterogeneous composites is a specific case of the modified theory.

As far as the types of inhomogeneity (classes of microstructures) are concerned, composite materials can conditionally be divided into ordered and random heterogeneous. Random heterogeneous composites include, for example, media with spherical, ellipsoidal or more complex-shaped inclusions, randomly dispersed in a matrix. In ordered structures (where inclusions are arranged in a strictly periodic manner), it is possible to obtain exact analytical expressions for the elastic moduli (Young's and shear moduli, Poisson's ratio etc.). With regard to the definition of the exact solution to the problem of calculating the effective moduli the reader is referred to Ref. [11]. The exact solution is required for obtaining in the closed form analytical expressions for the effective moduli, appropriate for an arbitrary large inhomogeneity of mechanical properties, e.g. for an arbitrary large ratio of Young's modulus of the first and second phases. Although in the real-world composites ideal periodic structures do not occur, analytical results for the elastic moduli of ordered structures are useful because they allow one to construct efficient computationally oriented models of multi-parameter complex systems by using asymptotic methods, which can, owing to their simplicity, be directly employed for control engineering of composite-material-based systems [12-14].

For random heterogeneous composite materials, it is impossible to obtain exact analytical solutions for the elastic moduli. In this case, sufficiently precise approximate expressions can be derived for small concentrations of one phase (isolated inclusions). Alternatively, good approximate solutions are possible in the percolation region, particularly in the so-called critical region and large inhomogeneity [1,2,15,16]. Both cases are interesting from the theoretical and experimental points of view. However, real-world composites usually do not belong to these two classes. Therefore, it is desirable to have an approximation, which describes the properties of a randomly inhomogeneous composite material in a broad concentration range, including large inhomogeneity.

For a significantly simpler (in comparison to the elasticity theory) problem of calculation of effective properties of electro-conductive (effective electrical conductivity coefficient) or dielectric (effective permittivity) composite materials, such an approximation is well known. It is the effective medium theory (EMT), based on the problem of an isolated inclusion and self-consistency considerations [1,2,15-17]. For the conductivity or permittivity problem, this estimation is often called the Bruggeman-Landauer (BL) approximation [18,19]. The BL approximation for the simplest two-phase isotropic case has the following form:

$$\frac{\sigma_e-\sigma_1}{2\sigma_e+\sigma_1}p+\frac{\sigma_e-\sigma_2}{2\sigma_e+\sigma_2}\left(1-p\right)=0, \qquad (1)$$

where $p$ is the concentration of the first phase, $\sigma_e$ is the effective electrical conductivity, $\sigma_1$ and $\sigma_2$ are the conductivities of the first and second phases, respectively. The percolation threshold in the BL approximation (1) for the large inhomogeneity ( $\sigma_1/\sigma_2 \to \infty$ ) is equal to 1/3. Simultaneously, one may calculate the effective elastic properties of the same composite. Rigorous cross-property relations linking the effective transverse electrical conductivity and the effective transverse elastic moduli of a fibre-reinforced (two-dimensional) composite have been investigated in Ref. [20]. Effective elastic properties of random two-dimensional composites have been written in analytic form with the accuracy of $O(p^4)$ in [21]. The classical EMT for the elasticity problem in the three-dimensional (3d) case has an inherent property that the percolation threshold is equal to 1/2 [22]. From our point view, there is a logical discrepancy, when simultaneous investigations of mechanical and electric properties contradict each other. In this context, it seems necessary to modify the EMT for elastic properties.

The paper is organized as follows: In Section 2, we briefly overview the conventional EMT approximation for the elasticity problem. In the following Section we begin with calculations of the percolation threshold, the Poisson's ratio at the threshold as well as the critical exponents. Next, we discuss the inconsistency of the conventional EMT for the elasticity problem in comparison with its application to the conductivity problem. To overcome this shortcoming of the theory, we propose a modification of the EMT, in which the percolation threshold can be prescribed. In Section 4, the percolation properties of the modified EMT are calculated and discussed. Finally, we consider the two-dimensional case, for which we succeed to construct the analog of the famous formula by A.M. Dykhne for the electrical conductivity of a symmetric two-phase composite [23] in the case of a symmetric elastic composite, using the modified theory. Conclusions are drawn in Section 5, where also an outlook into the future research is given.

**2. Materials and Methods**

In what follows, we consider random heterogeneous two-phase composites. The self-consistent EMT approximation considers inclusions of the spherical shape embedded into a fictitious homogeneous medium with the effective elastic properties searched. The concentration of the first phase is *p*, the concentration of the second phase is (1 - *p*). The mechanical properties of both phases and the effective medium are isotropic.

For the elasticity problem, the self-consistent EMT approximation was obtained in the past [1,5,6]. It can be written in the following form:

$$\left.\begin{array}{l} \dfrac{p}{1+\alpha_{\mathrm{e}}\left(\dfrac{K_1}{K_{\mathrm{e}}}-1\right)}+\dfrac{1-p}{1+\alpha_{\mathrm{e}}\left(\dfrac{K_2}{K_{\mathrm{e}}}-1\right)}=1 \\ \dfrac{p}{1+\beta_{\mathrm{e}}\left(\dfrac{G_1}{G_{\mathrm{e}}}-1\right)}+\dfrac{1-p}{1+\beta_{\mathrm{e}}\left(\dfrac{G_2}{G_{\mathrm{e}}}-1\right)}=1 \end{array}\right\}, \tag{2}$$

where

$$\alpha_{\mathrm{e}}=\frac{1}{3}\cdot\frac{1+\nu_{\mathrm{e}}}{1-\nu_{\mathrm{e}}},\quad \beta_{\mathrm{e}}=\frac{2}{15}\cdot\frac{4-5\nu_{\mathrm{e}}}{1-\nu_{\mathrm{e}}}, \tag{3}$$

and $G_{\mathrm{e}}, K_{\mathrm{e}}, \nu_{\mathrm{e}}$ are the effective shear modulus, the bulk modulus and the Poisson's ratio, respectively, while $G_1, G_2, K_1, K_2, \nu_1, \nu_2$ are the values of these moduli in the first and second phases.

For a small concentration of one phase ( $p \ll 1$), $\sigma_{\mathrm{e}}(p)$ from (1) and $G_{\mathrm{e}}(p)$ from (2,3) transform into the known expressions, obtained for small concentrations. The approximation (for the effective conductivity this is the Maxwell's approximation, see e.g. [1,2,15,16]) describes the behavior of the effective coefficients with the accuracy of the first power of concentration p. The Maxwell's approach can also be generalized to other physical properties, e.g. the piezoelectricity [24].

The most challenging for theoretical considerations region of filler concentration is around a particular concentration value, where with a small variation of the concentration, a significant change in the values of effective coefficients and moduli takes place at large inhomogeneity, see Figures 1 and 2. Such a concentration is denoted the percolation threshold. According to the percolation theory in a vicinity of the percolation threshold (in the so-called critical region, $|p-p_c|<<1$) the effective conductivity ($\sigma_e$), shear ($G_e$) and Young's ($E_e$) moduli have the following power-law behavior:

$$\sigma_e \sim (p-p_c)^t,\ \sigma_2=0,\quad p>p_c;\qquad \sigma_e \sim (p_c-p)^{-q},\ \sigma_1=\infty,\quad p<p_c; \tag{4}$$

$$E_e \sim (p-p_c)^f,\ G_e \sim (p-p_c)^f,\ E_2=G_2=0,\ p>p_c; \tag{5}$$

$$E_e \sim (p_c-p)^{-S},\ G_e \sim (p_c-p)^{-S},\ E_1=G_1=\infty,\ p<p_c, \tag{6}$$

where *f* and *S* are universal critical exponents [1,2,15,16].

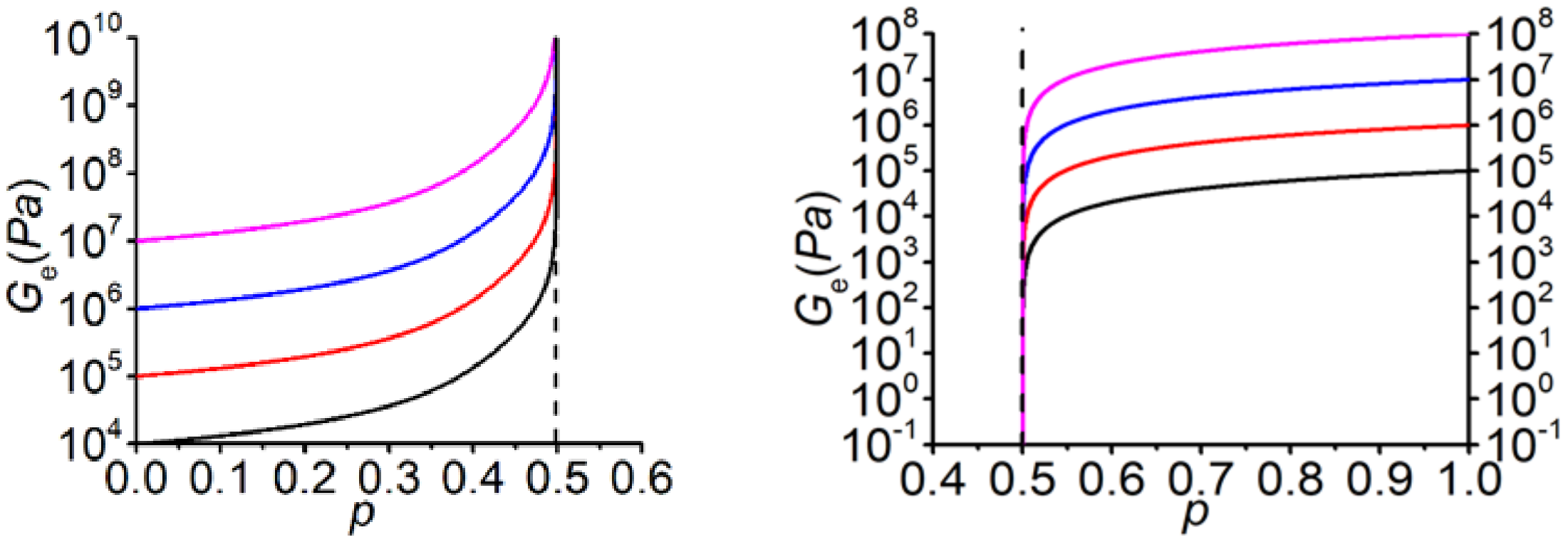

(**a**) (**b**)

**Figure 1.** Concentration dependences of the effective shear modulus $G_e$ at $G_2/G_1=0$. (**a**) Different shear moduli of the second phase $G_2$ ($\nu_2=0.49$) and $G_1=\infty$. (**b**) Different shear moduli of the first phase $G_1$ ($\nu_1=0.3$) and $G_2=0$ Pa.

Approximate values of critical exponents can be obtained from qualitative considerations, while their precise values can be derived from numerical modeling [15,22,25].

The EMT for the elasticity problem approximately describes the behavior of elastic moduli in the critical region. Note that the behavior of effective conductivity and elastic moduli at large inhomogeneity is analogous to the behavior of the order parameter in the theory of second-order phase transitions [26]. Calculation of the critical exponents of the order parameter in the EMT gives their approximation value. The clarification of these values is more complicated and related to the consideration of fluctuations of the order parameter [27-29]. Similarly, EMT in the elasticity problem yields approximate values of critical exponents *f* and *S*. According to the EMT, the critical exponents of the effective conductivity are easily found from (1) and they are equal to unity: $t=q=1$ [1,2]. According to experiments and numerical calculations $t=2$, $q=0.73$ in the three-dimensional case [15].

In spite of its seeming simplicity, the BL approximation, based on the idea of a self-consistent solution, describes the variation of effective coefficients in the entire concentration quantitatively and qualitatively. For example, it could be the effective electrical conductivity $\sigma_e$ (1). At small inhomogeneity ($\sigma_1/\sigma_2$ is slightly larger than unity), for a well homogenized mixture of two phases with conductivities $\sigma_1$ and $\sigma_2$, the dependence $\sigma_e(p)$, obtained from (1), describes the experimental data well. As is well known [1,2,15,16], at large inhomogeneity ($\sigma_1/\sigma_2 \gg 1$) there exist such a value of concentration, called the percolation threshold $p_c$, in the vicinity of which the effective conductivity (other effective coefficients and moduli as well) steeply change changes its value. The description of such phenomena is the subject of the percolation theory, which is the analogue of the theory of the second-order phase transitions. The main geometrical element there is the connectivity through one of the phases - the appearance of a continuous path via one of the phases through the entire specimen (the so-called infinite percolation cluster). At a first glance, an EMT approximation (calculation of the field inside the isolated spherical inclusion of one of the phases, immersed into the effective medium) cannot provide a description for the steep rise of the effective conductivity $\sigma_e$. However, the BL approximation, not only gives a growth in the vicinity of a some concentration value at $\sigma_1/\sigma_2 \gg 1$, but also gives a power-law dependence of $\sigma_e$ on $|p - p_c|$ in a vicinity of $p_c$ at $\sigma_1 = \infty$ , $\sigma_2 \neq \infty$ or at $\sigma_1 \neq 0$ , $\sigma_2 = 0$ . Such power-law dependences close to the percolation threshold take place for the order parameters (in the theory of the second-order phase transitions) and for the effective coefficients and moduli in the percolation theory (4-6).

Thus, the EMT approximation qualitatively and partly also quantitatively describes the percolation behavior. Therefore, it is reasonable to use the notations of the percolation theory – percolation threshold, critical concentration region ($|p - p_c| \ll 1$) as well as critical exponents describing the power-law behavior in the critical region. For the conductivity problem, the EMT approximation leads to the existence of the percolation threshold $p_c^\sigma$ (the upper index denotes the conductivity problem) and the power-law behavior of the conductivity $\sigma_e$ in the vicinity of $p_c^\sigma$, characterized by its own critical exponents. In the EMT framework, $p_c^\sigma = 1/3$.

It is worth noting that the EMT results, when describing the efficient coefficients and moduli are approximate, similarly to the Landau's theory of the second order phase transition used for the description of the order parameters.

## 3. Results

### *3.1. Percolation threshold in traditional EMT*

Let us first determine the percolation threshold $p_c$. We reformulate the EMT equations (2), using as the independent variables "$G$"s and "$\nu$"s. Taking into account the known relation $K = 2G(1+\nu)/3(1-2\nu)$, the system of equations (2) can be written in the form:

$$\left.\begin{aligned} &\frac{p}{1+\alpha_e\left(\dfrac{G_1}{G_e}\dfrac{1+\nu_1}{1+\nu_e}\dfrac{1-2\nu_e}{1-2\nu_1}-1\right)}+\frac{1-p}{1+\alpha_e\left(\dfrac{G_2}{G_e}\dfrac{1+\nu_2}{1+\nu_e}\dfrac{1-2\nu_e}{1-2\nu_2}-1\right)}=1 \\ &\frac{p}{1+\beta_e\left(\dfrac{G_1}{G_e}-1\right)}+\frac{1-p}{1+\beta_e\left(\dfrac{G_2}{G_e}-1\right)}=1 \end{aligned}\right\} \quad (7)$$

If a composite with $E_2 = G_2 = 0$ is considered, taking in (7) $G_2 = 0$ it is possible to investigate the case when $p \to p_c$, $G_e \to 0$. Equations (7) are simplified and can be written in the following form

$$\left.\begin{aligned}&\left[\frac{1}{1+\alpha_{\mathrm{e}}\left(\dfrac{G_1}{G_{\mathrm{e}}}\cdot\dfrac{1+\nu_1}{1+\nu_{\mathrm{e}}}\cdot\dfrac{1-2\nu_{\mathrm{e}}}{1-2\nu_1}-1\right)}-1\right]p+\left(\frac{1}{1-\alpha_{\mathrm{e}}}-1\right)(1-p)=0\\&\left[\frac{1}{1+\beta_{\mathrm{e}}\left(\dfrac{G_1}{G_{\mathrm{e}}}-1\right)}-1\right]p+\left(\frac{1}{1-\beta_{\mathrm{e}}}-1\right)(1-p)=0.\end{aligned}\right\}. \quad (8)$$

Putting now $G_{\mathrm{e}}\to 0$ and $p\to p_{\mathrm{c}}$, from (8) we find

$$\left.\begin{aligned}&-p_{\mathrm{c}}+\left(\frac{1}{1-\alpha_{\mathrm{e}}}-1\right)(1-p_c)=0\\&-p_{\mathrm{c}}+\left(\frac{1}{1-\beta_{\mathrm{e}}}-1\right)(1-p_c)=0\end{aligned}\right\}. \quad (9)$$

From (9) we obtain

$$\alpha_{\mathrm{e}}=\beta_{\mathrm{e}}. \quad (10)$$

It immediately follows that

$$\nu_{\mathrm{e}}\left(p=p_{\mathrm{c}}^{\mathrm{E}}\right)=\frac{1}{5}, \qquad p_{\mathrm{c}}^{\mathrm{E}}=\frac{1}{2}, \quad (11)$$

where the upper index of $p_{\mathrm{c}}^{E}$ means that this percolation threshold refers to the elasticity problem.

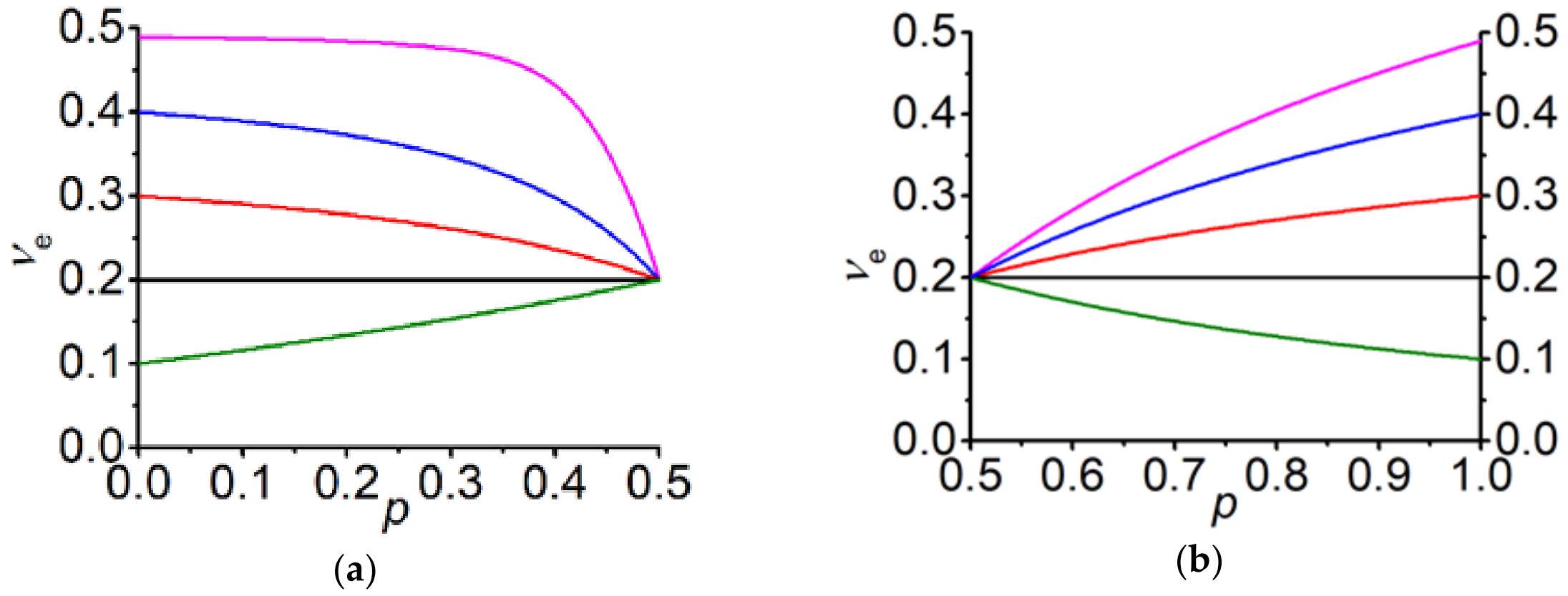

**Figure 2.** Concentration dependences of the effective Poisson's ratio $\nu_{\mathrm{e}}$ at $G_2/G_1=0$. (**a**) Different Poisson's ratios the second phase $\nu_2$ ($G_2=10^5$ Pa) and $G_1=\infty$. (**b**) Different Poisson's ratios of the first phase $\nu_1$ ($G_1=10^{11}$ Pa) and $G_2=0$ Pa.

Note that the value of the Poisson's ratio $\nu_{\mathrm{e}}$ at the percolation threshold (10) in the EMT framework doesn't depend on the mechanical parameters of phases, i.e. on $\nu_1,\nu_2$ and $G_1$: $\nu_{\mathrm{e}}\left(p\to p_{\mathrm{c}}^{E}=1/2\right)=1/5$. The case $G_2/G_1\to 0$ is considered above at $G_2\to 0$ for a finite value of $G_1$. The same result also takes place at $G_2/G_1\to 0$, but for a non-zero value of $G_2$, while $G_1\to\infty$ see Appendix A. The particular value $\nu^*=1/5$ was discussed in [30]. Later we will return to the discussion of the particular values of the Poisson's ratio. Figure 2 shows the concentration behavior of the effective Poisson's ratio for the both cases.

Figure 3 demonstrates the concentration dependences of the effective shear modulus and the effective Poisson's ratio for the finite values of the shear modulus of both constitutive materials. The inset shows enlarged the region in a vicinity of the percolation threshold. It is seen that the effective Poisson's ratio is not exactly equal to 1/5 at $p$ = 1/2, as it has been shown previously in [30].

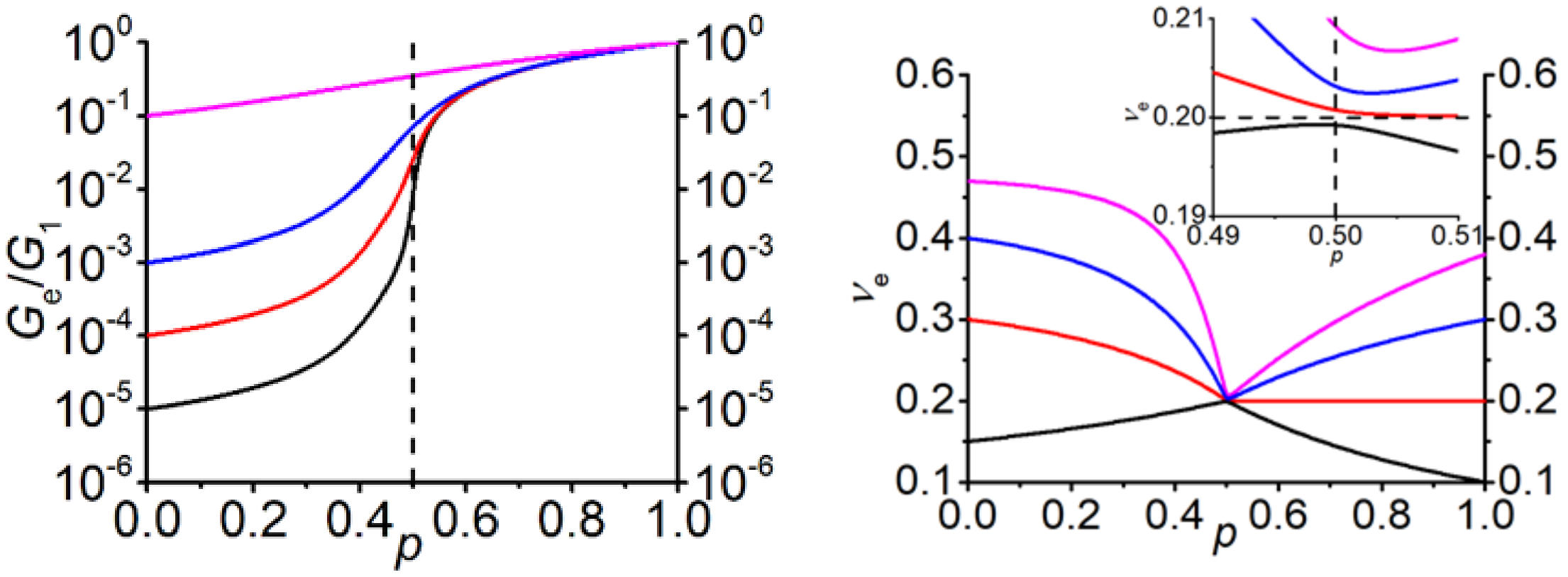

(**a**) (**b**)

**Figure 3.** (**a**) Concentration dependence of the normalized effective shear modulus $G_e/G_1$ for different shear moduli of the second phase. $G_1 = 10^{11}\,\text{Pa}$, $\nu_1 = 0.3$, $\nu_2 = 0.49$; (**b**) Concentration dependence of the effective Poisson's ratio for different combinations of the Poisson's moduli of the phases. $G_1 = 10^{11}\,\text{Pa}$. The inset shows enlarged the region around the percolation threshold.

*3.2 Critical exponents in traditional EMT*

By using the known value of the percolation threshold $p_c^E = 1/2$, it is possible to decompose the expression for $G_e$ into the series with respect to the vicinity to the percolation threshold $|p - p_c|$ and calculate the corresponding critical exponents.

At $G_2 \to 0$ we obtain from the first equation of system (7)

$$\nu_e = \frac{G_1\left[(1-\nu_1) - 3p(1+\nu_1)\right] + 2G_e(1-2\nu_1)}{G_1\left[(1+\nu_1) + 3p(1+\nu_1)\right] + 2G_e(1-2\nu_1)} \tag{12}$$

and from the second equation of system (7) we get two solutions:

$$\nu_e = \frac{4}{5}, \qquad \nu_e = \frac{1}{5}\cdot\frac{(8-15p)G_1 + 7G_e}{(2-3p)G_1 + G_e}. \tag{13}$$

The first solution in (13) has no physical meaning, because the Poisson's ratio is always not larger than 1/2.

Introducing a variable $\tilde{G}_e = G_e/G_1$, and equating the right-hand sides of obtained solutions (12,13):

$$\frac{1}{5}\cdot\frac{8-15p+7\tilde{G}}{2-3p+\tilde{G}}=\frac{(1-\nu_1)-3p(1+\nu_1)+2\tilde{G}_e(1-2\nu_1)}{(3p+1)(1+\nu_1)+2\tilde{G}_e(1-2\nu_1)}. \tag{14}$$

After cross-multiplication we obtain a quadratic equation for $\tilde{G}_e$ with two solutions:

$$\tilde{G}_e=\frac{-10\nu_1-3p(3-7p\nu_1)+8\pm\sqrt{\omega}}{8\left(2-\nu_1\right)}, \tag{15}$$

where

$$\omega=9(7\nu_1-3)^2p^2-12(51\nu_1^2-35\nu_1+4)p+4(7\nu_1-2)^2. \tag{16}$$

The solution with the plus sign in front of the root term in the numerator of (15) does not possess the necessary physical properties ($G_e\to 0$ at $p\to p_c$).

Near the percolation threshold, $\tilde{G}_e$ behaves in a power-law manner (5), as it should be expected in the critical region of the percolation theory. Substituting (16) into (15), we find the critical exponent f by the method of Padé approximants [31, 32]

$$f_{\text{EMT}}=\lim_{p\to 1/2^+}\left\{\left(p-\frac{1}{2}\right)\left[\frac{\partial}{\partial p}\ln\left(\tilde{G}_e\right)\right]\right\}=1 \tag{17}$$

where the notation $f_{\text{EMT}}$ signifies that this value is obtained using the EMT and the result coincides with the value obtained in [22]. The value obtained by numerical calculations outside of the EMT framework is $f$ = 3.76 [1,2,15,16,25]. A proof for $G_1\to\infty$ is given in Appendix B.

The availability of percolation thresholds at large inhomogeneity, both for the conductivity problem as well as the elasticity problem, is, of course, related to the formation in a two-phase randomly inhomogeneous system of the so-called infinite cluster [1,2,15,16], which is a connected path through the entire specimen *via* one of the phases. In the case of the conductivity problem, the appearance of an infinite cluster of the first phase ($\sigma_1\gg\sigma_2$) means the appearance of a well conducting path, and therefore, a steep decline in the electrical resistance of the specimen. For the elasticity problem, at $E_1\gg E_2$ that means the appearance of a rigid frame. Close to the concentration, where the formation of "percolation" takes place (i.e. an infinite cluster of one of the phases comes into existence) the power-law dependence of the effective coefficients is observable.

*3.3 Criticism of the traditional EMT for the elasticity problem*

In spite of the fact that effective coefficients and moduli give correct qualitative, and in some cases even quantitative, behavior in the entire concentration range, both for conductivity and elasticity calculations, there exist principle difficulties and logical discrepancies in the conventional approach.

The main disadvantage is the strict equality of $p_c^E$ to 1/2. EMT describes random heterogeneous media, in other words, well-blended mixtures of two phases. At small concentration of the first phase, $p\ll 1$, such a medium represents the inclusions of the first phase in the second phase. At the concentration $p$ close to 1, these are the inclusions of the second phase in the first phase. There is a symmetry: mutual interchange of phases $1\rightleftarrows 2$ and phases $p\rightleftarrows 1-p$ does not change the effective properties of such symmetric composites. This symmetry is directly seen in (1) for the conductivity problem and in (2,3) for the elasticity problem. For example, after replacing in (2,3) $G_1\rightleftarrows G_2$, $K_1\rightleftarrows K_2$, $p\rightleftarrows 1-p$ the system of equation remains unaltered [33, 34]. If the critical concentration of the first phase is equal to $p_{c1}$, by virtue of said symmetry, the percolation threshold $p_{c2}$ of the second phase is equal to $\left(1-p_{c1}\right)$, because of the parity of the phases.

If the concentration, e.g. of the first phase, is increased, upon reaching $p=p_c$ in a three-dimensional medium, an infinite cluster of the first phase arises. At the same time, an infinite cluster of the second phase doesn't disappear. According to the symmetry the cluster of the second phase emerges at $1-p_c$ (while decreasing from 1). Thus, at $p_c<p<1-p_c$, there exist two clusters in the medium simultaneously. In the two-dimensional case, the percolation threshold of the first phase $p_c$ has the value of 1/2 in the EMT-approximation and upon occurrence of the cluster of one phase the second cluster disappears. For the two-dimensional (2d) case, the percolation threshold for the first phase in the EMT is 1/2 and, due to the symmetry, the percolation threshold of the second phase is also 1/2, thus they coincide. Indeed, in the 2d symmetrical case the presence of an infinite cluster of one phase excludes the existence of an infinite cluster of another phase. In the three-dimensional case, with the existence of symmetry, this is not true. Therefore, $p_c^E=1/2$ seems to be at least peculiar in the three-dimensional case, because the EMT describes a random heterogeneous medium.

One more disadvantage of the EMT-approximation is a mismatch between thresholds $p_c^\sigma$ and $p_c^E$. Of course, the problems of conductivity and elasticity represent different, from the point of view of mathematical physics, equations (even of different order, see the elasticity theory [35, 36]). However, at the same time EMT for conductivity should not contradict to the EMT for elasticity. For example, if the effective conductivity (dielectric permittivity) and effective elastic moduli are measured simultaneously, their steep change should occur at the same concentration. That concentration, at which an infinite cluster arises, see, e.g. experimental data [37, 38], where for the two- and three-dimensional cases it was experimentally shown that $p_c^\sigma=p_c^E$.

For the elimination of the inconsistency between the conductivity and elasticity problems, an EMT modification is required, which allows one to prescribe a preselected percolation threshold and, therefore, to match them in both problems.

Depending on the fabrication technology (and thereby the presence of different correlations in the location of inclusions), the numerical value of the percolation threshold varies [33]. For example, for the case of porous materials it is well known that the percolation threshold depends on the particular material composition [39, 40]. At the same time, the EMT for conductivity (the BL approximation) gives a threshold value $p_c^\sigma=1/3$, which does not allow one to describe the experimental data in all cases. In the work [41], the EMT for galvanomagnetic phenomena (electrical conductivity in a magnetic field) has been modified in such a way that it became possible to set the percolation threshold. This means it is possible to obtain the field and concentration dependences of the effective components of the conductivity tensor in the framework of EMT with different percolation thresholds. Such a modified theory made it possible to explain the results of many previous experiments, which seemed to be paradoxical in that time, see for example, [42-48]. For a specific case, when the effective conductivity tensor is a scalar function of conductivity, the modified EMT approximation can be written as:

$$\frac{\dfrac{\sigma_e-\sigma_1}{2\sigma_e+\sigma_1}}{1+c(p,\tilde{p}_c)\dfrac{\sigma_e-\sigma_1}{2\sigma_e+\sigma_1}}p+\frac{\dfrac{\sigma_e-\sigma_2}{2\sigma_e+\sigma_2}}{1+c(p,\tilde{p}_c)\dfrac{\sigma_e-\sigma_2}{2\sigma_e+\sigma_2}}\left(1-p\right)=0, \quad (18)$$

where $c(p,\tilde{p}_c)$ in the following will be called the SV term (owing to Sarychev and Vinogradov [41]). It has the following form:

$$c(p,\tilde{p}_{\rm c})=(1-3\tilde{p}_{\rm c})\left(\frac{p}{\tilde{p}_{\rm c}}\right)^{\tilde{p}_{\rm c}}\left(\frac{1-p}{1-\tilde{p}_{\rm c}}\right)^{1-\tilde{p}_{\rm c}} \tag{19}$$

and $\tilde{p}_{\rm c}$ is the preselected percolation threshold.

From the modified EMT (18,19), the same critical behavior is obtained in the vicinity of the percolation threshold as in the conventional EMT, but at the prescribed value of the percolation threshold. In Ref. [49] we introduced the method of the moveable (field-dependent) percolation threshold and employed the modified EMT (18)-(19) for describing the significant changes of dielectric and magnetic properties of the magnetoactive elastomers in an external magnetic field.

### *3.4 Modification of the effective medium theory for the elasticity problem*

According to the aforementioned consideration (a discrepancy between the percolation thresholds in the conductivity and elasticity calculations) it becomes clear that the standard EMT for the elastic properties (2,3) has to be modified. From our point of view, the remedy could be to do in the same manner as for the conductivity calculation in the EMT, namely it should be possible to preselect the percolation threshold. Although the physical processes of electrical conductivity or elasticity and the equations describing them are different, the EMT approximation for calculating the effective coefficients (moduli) (2,3) can be modified in a similar way. For convenience, we write the EMT approximation for elastic effective modules (2,3) in the following form (*cf.* the EMT approximation for conductivity (1))

$$\left.\begin{array}{l}\Omega_1 p+\Omega_2(1-p)=0\\ \Theta_1 p+\Theta_2(1-p)=0\end{array}\right\}, \tag{20}$$

where

$$\Omega_{\rm i}=\frac{\dfrac{G_{\rm i}}{G_{\rm e}}\cdot\dfrac{1+\nu_{\rm i}}{1+\nu_{\rm e}}\cdot\dfrac{1-2\nu_{\rm e}}{1-2\nu_{\rm i}}-1}{1+\alpha_{\rm e}\left(\dfrac{G_{\rm i}}{G_{\rm e}}\cdot\dfrac{1+\nu_{\rm i}}{1+\nu_{\rm e}}\cdot\dfrac{1-2\nu_{\rm e}}{1-2\nu_{\rm i}}-1\right)},\qquad \Theta_{\rm i}=\frac{\dfrac{G_{\rm i}}{G_{\rm e}}-1}{1+\beta_{\rm e}\left(\dfrac{G_{\rm i}}{G_{\rm e}}-1\right)},\qquad {\rm i}=1,2. \tag{21}$$

Recall that $\alpha_{\rm e}$ and $\beta_{\rm e}$ for the three-dimensional case are given in (3).

In this paper, we propose a modified EMT for the elasticity problem, replacing (20) with the following system of equations

$$\left.\begin{array}{l}\dfrac{\Omega_1}{1+s(p,\tilde{p}_{\rm c})\Omega_1}p+\dfrac{\Omega_2}{1+s(p,\tilde{p}_{\rm c})\Omega_2}(1-p)=0\\ \dfrac{\Theta_1}{1+s(p,\tilde{p}_{\rm c})\Theta_1}p+\dfrac{\Theta_2}{1+s(p,\tilde{p}_{\rm c})\Theta_2}(1-p)=0\end{array}\right\}, \tag{22}$$

where the introduced by us term s is similar, but not equal to the SV term (19)

$$s(p,\tilde{p}_{\rm c})=(1-2\tilde{p}_{\rm c})\left(\frac{p}{\tilde{p}_{\rm c}}\right)^{\tilde{p}_{\rm c}}\left(\frac{1-p}{1-\tilde{p}_{\rm c}}\right)^{1-\tilde{p}_{\rm c}} \tag{23}$$

In the case when $\tilde{p}_{\rm c}=p_{\rm c}^E=1/2$, the term $s(p,\tilde{p}_{\rm c}=p_{\rm c}^E)$ vanishes and (22) goes into the conventional EMT for the elasticity problem (20).

Now, by preselecting $p_{\rm c}$ in the term $s(p,\tilde{p}_{\rm c})$ (23) one can find the concentration behavior $G_{\rm e}(p)$, which, as it should be, has the given percolation threshold, see Figures 4 and 5. Taking into account that $E_e = 2G_e(1+\nu_e)$, the effective Young's modulus has the same percolation threshold as the shear modulus $G_e$.

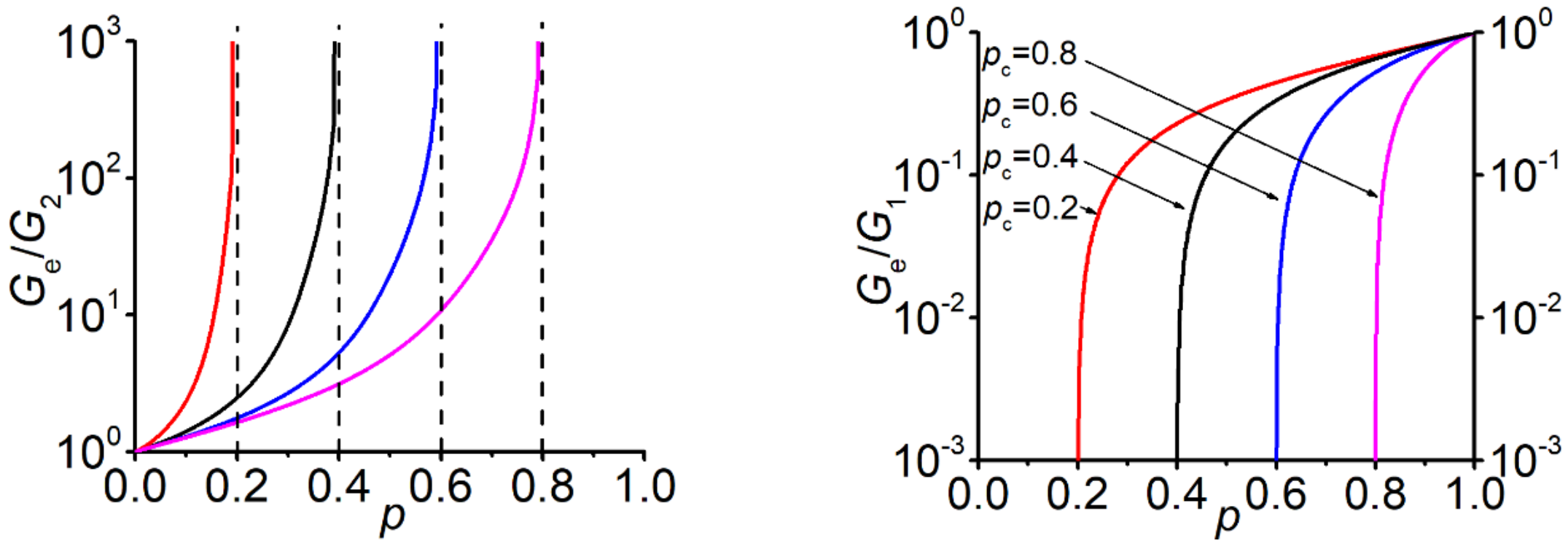

(**a**) (**b**)

**Figure 4.** Concentration dependences of the effective shear modulus $G_e$ at $G_2/G_1 = 0$ for various values of the percolation threshold. (**a**) $G_2 = 10^5\,\mathrm{Pa}$, $\nu_2 = 0.49$ and $G_1 = \infty$. (**b**) $G_1 = 10^{11}\,\mathrm{Pa}$, $\nu_1 = 0.3$ and $G_2 = 0$ Pa.

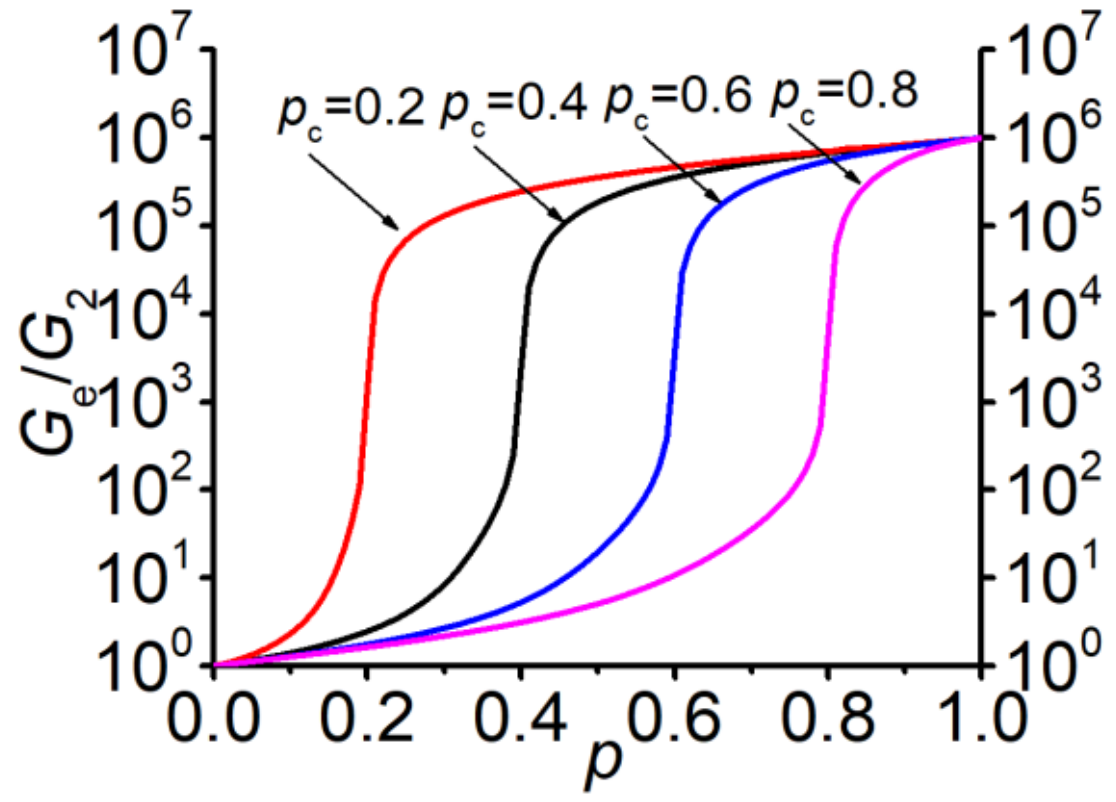

**Figure 5.** Concentration dependence of the normalized effective shear modulus $G_e/G_2$ for various values of the percolation threshold. $G_1 = 10^{11}\,\mathrm{Pa}$, $\nu_1 = 0.3$, $G_2 = 10^5\,\mathrm{Pa}$, $\nu_2 = 0.49$.

In contrast to the previous, multi-parametric theoretical approaches or cascade continuum micromechanics models (cf., e.g. [40]), our model uses only the most important characteristic – the percolation threshold, which is determined by the fabrication method of a composite material.

## 4. Discussion

### *4.1. Percolation properties of the modified EMT*

Let us consider the behavior of the elastic moduli in the modified EMT.

#### 4.1.1 Calculation of the percolation threshold

Let us first prove that $p_c$ is indeed the percolation threshold of equations (22,23). Let us set in (23) $p_c$ equal to $\tilde{p}_c$. Taking $G_2 = 0$ and considering $G_e \to 0$ at $p \to p_c$ we obtain:

$$\left.\begin{aligned}(1-p-s(p,\tilde{p}_{c}))\left(\frac{1}{1-\alpha_{e}}-1\right)=p\\(1-p-s(p,\tilde{p}_{c}))\left(\frac{1}{1-\beta_{e}}-1\right)=p\end{aligned}\right\}. \qquad (24)$$

Taking into account $p = p_c$ we get

$$\alpha_e = \beta_e = 1/2. \qquad (25)$$

Then, the multipliers in the large round brackets in (24) are equal to 1 and we obtain from (24), taking into account the expression for $s(p,\tilde{p}_c)$ (23),

$$2p_c = 1-(1-2\tilde{p}_c)\left(\frac{p}{\tilde{p}_c}\right)^{\tilde{p}_c}\left(\frac{1-p}{1-\tilde{p}_c}\right)^{1-\tilde{p}_c}\Bigg|_{p=p_c}, \qquad (26)$$

where $\tilde{p}_c$ is the preselected percolation threshold, while $p_c$ is the percolation threshold resulting from the EMT (24). It is easily seen that the solution of Equation (26) with respect to $p_c$ is

$$p_c = \tilde{p}_c \qquad (27)$$

i.e. the percolation threshold $p_c$ is indeed equal to the given value $\tilde{p}_c$, see Figure 5. Similar proof can be provided for $G_1 \to \infty$, see Appendix C.

4.1.2 Effective Poisson's ratio

Figure 6 presents the concentration behavior of the effective Poisson's ratio $\nu_e$ for various percolation thresholds and finite values of $G_1$ and $G_2$.

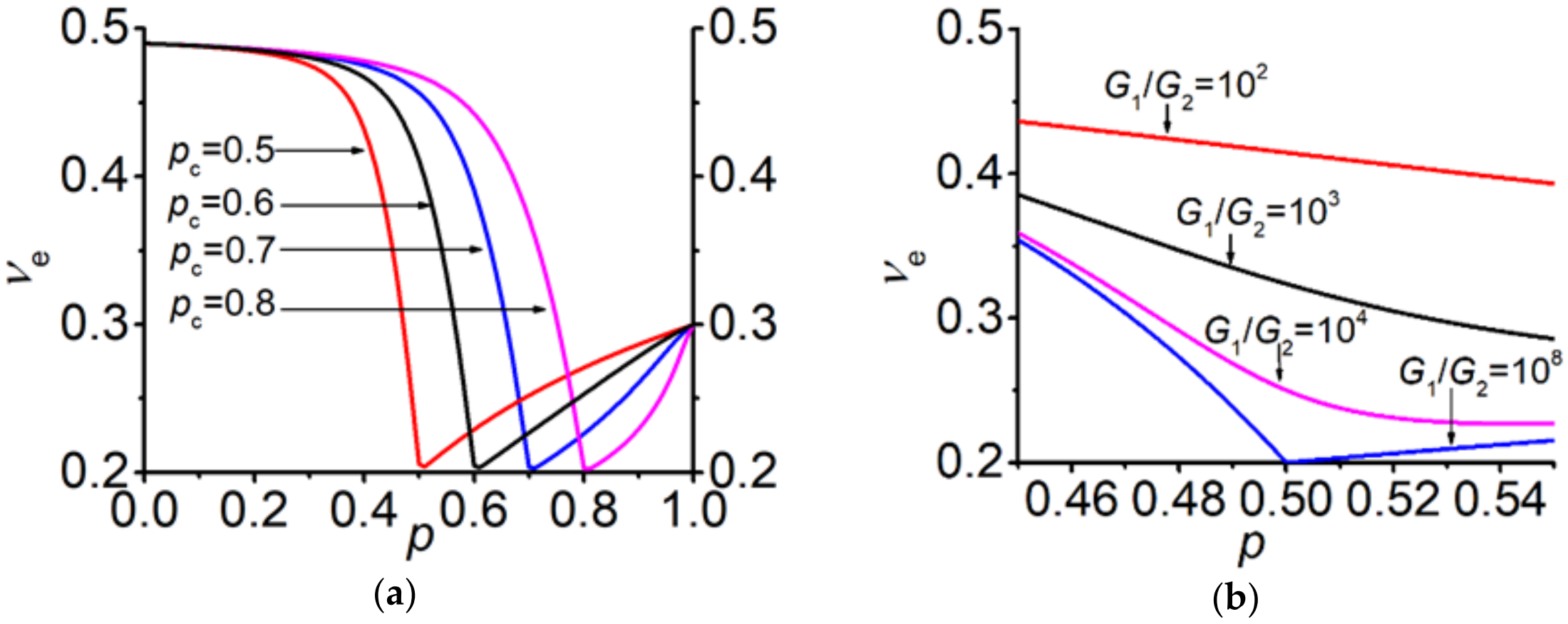

**Figure 6.** Concentration dependences of the effective Poisson's ratio $\nu_e$. (**a**) Variation of the percolation threshold. $G_1 = 10^{11}$ Pa, $\nu_1 = 0.3$, $G_2 = 10^5$ Pa, $\nu_2 = 0.49$; (**b**) Variation of the ratio $G_1/G_2$ at $p_c = 1/2$. $\nu_1 = 0.3$, $\nu_2 = 0.49$.

Taking into account (25), we find that at the percolation threshold (which we now defined ourselves)

$$\nu_{\mathrm{e}}\left(p=\tilde{p}_{\mathrm{c}}, G_2=0\right)=\nu^*=1/5. \tag{28}$$

Note that in the case when $G_1 \to \infty$, while $G_2$ remains finite ($G_1 / G_2 \to \infty$), the equality (28) also holds $\nu_{\mathrm{e}}\left(p=\tilde{p}_{\mathrm{c}}, G_1=\infty\right)=\nu^*=1/5$. It is well known that behavior of the effective Poisson's ratio is somewhat anomalous, for example, it does not necessarily lie between the Poisson's ratio of the matrix and inclusion phases [50]. In [30], it was pointed out that there is a specific value of the Poisson's ratio $\nu^*$, which is equal to $1/5$ in the three-dimensional case. It was shown that, if $\nu_1=\nu_2 \le \nu^*$ then for any ratio of $G_2 / G_1$ the inequality $\nu_{\mathrm{e}} \le \nu^*$ holds. Vice versa, if $\nu_1=\nu_2 \ge \nu^*$ then $\nu_{\mathrm{e}} \ge \nu^*$ is fulfilled. Note that the value $\nu^*$ is achieved at $p=p_c$ and only with the infinite heterogeneity, while with a finite ratio of $G_1 / G_2$, the value $\nu_{\mathrm{e}}$ tends to, but does not reach this value. The larger the ratio $G_1 / G_2$, the closer $\nu_{\mathrm{e}}$ comes to $\nu^*$, see Figure 6b. An exception is the case when the values of the Poisson's ratio in both phases are equal to $\nu^*$. Then $\nu_{\mathrm{e}}=\nu^*$ does not depend on the concentration at any ratio of $G_1 / G_2$. Experimentally, the decrease of the Poisson's ratio with the decreasing foam density was observed in the metallic foams, where the Poisson's ratio approached the value of 0.21 [51].

#### 4.1.3 Critical exponents

For the case of modified equations (22)-(23), we did not succeed in calculating the critical exponents analytically, as it has been done in (15)-(17). However, the critical exponents *f* and *S* can be easily obtained numerically, as it is explained in Appendix B. It turns out that the SV modification of the EMT does not change the values of both exponents: *f* = *S* = 1.

### *4.2. The two-dimensional case*

Let us consider briefly the two-dimensional (2d) problem. In the 2d case, equations (1) remain the same but the parameters in them take the form of

$$\alpha_{\mathrm{e}}=\frac{1+\nu_{\mathrm{e}}}{2}, \quad \beta_{\mathrm{e}}=\frac{3-\nu_{\mathrm{e}}}{4}. \tag{29}$$

As in the previous case, it can be shown, that the percolation threshold in the 2d case is

$$p_{\mathrm{c}}^E\left(d=2\right)=2/3. \tag{30}$$

where *d* is the dimensionality of the problem [30]. This result contradicts the general considerations of a two-dimensional random heterogeneous medium.

The geometric structure of the mutual arrangement of phases does not depend on the type of physical processes investigated on this structure. In particular, for the two-dimensional conductivity problem in a random heterogeneous medium with the conductivity of phases $\sigma_1$ and $\sigma_2$, the exact result by A.M. Dykhne is known at the percolation threshold (for an arbitrary ratio of phase conductivities) [23]

$$\sigma_{\mathrm{e}}=\sqrt{\sigma_1\sigma_2}, \quad p=p_{\mathrm{c}}=1/2, \tag{31}$$

where, of course, the effective conductivity is symmetric with respect to the interchange of phase values, $\sigma_1 \rightleftarrows \sigma_2$.

For the elasticity problem at the percolation threshold (30), the effective elasticity moduli, contrary to the general considerations, are not symmetrical when the values of the modules are interchanged.

As in the three-dimensional case, we propose a modified mean-field theory for the elasticity problem. The form of the modified equations is the same as in the three-dimensional case (22), but in

(21) it is necessary to substitute the corresponding values of $\alpha_e$ and $\beta_e$ (29) and to write down the correction $s\left(p,\tilde{p}_c\right)$ for the two-dimensional case

$$s\left(p,\tilde{p}_{\rm c}\right)=(1-\frac{3}{2}\tilde{p}_{\rm c})\left(\frac{p}{\tilde{p}_{\rm c}}\right)^{\tilde{p}_{\rm c}}\left(\frac{1-p}{1-\tilde{p}_{\rm c}}\right)^{1-\tilde{p}_{\rm c}}. \qquad (32)$$

It is worth noting that, in the case of *d* dimensions, the percolation threshold in the traditional EMT for the elastic properties $p_{\rm c}^E$ has the following value: $p_{\rm c}^E=2/(d+1)$ [22]. In the *d*-dimensional case, the correction term $s\left(p,\tilde{p}_c\right)$ can be written as

$$s\left(p,\tilde{p}_{\rm c}\right)=\left(1-\frac{\tilde{p}_{\rm c}}{p_{\rm c}^E}\right)\left(\frac{p}{\tilde{p}_{\rm c}}\right)^{\tilde{p}_{\rm c}}\left(\frac{1-p}{1-\tilde{p}_{\rm c}}\right)^{1-\tilde{p}_{\rm c}}. \qquad (33)$$

For a random heterogeneous inhomogeneous medium $p_{\rm c}=\tilde{p}_{\rm c}=1/2$ and (32) takes the following form

$$s\left(p,\tilde{p}_{\rm c}=1/2\right)=\frac{1}{2}\sqrt{p\left(1-p\right)}. \qquad (34)$$

Note that the statements about the behavior of the effective Poisson module in the case of inequalities $\nu_1=\nu_2\le\nu^*$ or $\nu_1=\nu_2\ge\nu^*$ remain valid for the two-dimensional case [30] as well as in the modified theory, taking into account that in the two-dimensional case $\nu^*=1/3$.

It is easy to show that for the particular case $\nu_1=\nu_2=\nu^*$, the effective value of the Poisson's ratio does not depend on the concentration and values of the shear moduli $G_1$ and $G_2$, namely $\nu_{\rm e}=\nu^*=1/3$. Using this value, we can obtain an expression for the shear and Young's moduli at the percolation threshold, an analog of the Dykhne's expression (31).

$$G_{\rm e}=\sqrt{G_1G_2}, \quad E_{\rm e}=\sqrt{E_1E_2}, \quad \nu_1=\nu_2=\nu^*, \quad d=2. \qquad (35)$$

*4.3. Comparison with experiments*

The 2d experiments were reported, e.g. in [37,52]. In [37], the same percolation threshold $p_{\rm c}=0.6$ was found for the conductivity and elasticity phenomena. Similar measurements of the mechanical stiffness and electrical conductance were performed in [52], were the same percolation threshold $p_{\rm c}=0.331$ was determined for both physical properties. In both experiments, the ratio $G_1/G_2$ was equal to zero, and, therefore, the percolation threshold was well pronounced. Notice that the measured values of the percolation significantly differ from the percolation threshold in the traditional EMT $p_{\rm c}^E\left(d=2\right)=2/3$ and emphasize the significance of our considerations. Further, a three-dimensional percolation system was investigated in [38], where it was also found that electrical conductivity and Young's modulus have one and the same percolation threshold, which was not equal to 1/2.

In [53], the percolation transition and the elastic properties of block copolymers of styrene and butadiene were investigated (3d case). The ratio of elastic moduli was finite: $G_1/G_2=581.82$. As can be seen from Figure 3 (a), for a finite ratio of $G_1/G_2\ne\infty$, the percolation transition is not clearly pronounced in the concentration dependence of the effective elastic modulus. The particular concentration value, where the maximum change in behavior of the effective modulus occurs, is shifted towards concentrations smaller than 1/2. Note that analogous situation occurs for the effective electrical conductivity, which has been studied in detail in [49]. Figure 7 compares the results of the EMT with experimental results of [53]. A good agreement between the experimental results and the EMT has be achieved for the value close to that of the traditional EMT: $\tilde{p}_c \approx 0.50$. This result contradicts the percolation threshold value of 0.4 determined in [53] and the value of $p_c \approx 0.2085$ obtained for this data in [54] within the framework of a fractal model using the iterative averaging approach. We believe that these deviations are caused by the finite ratio of the shear moduli of the constitutive components, which should be taken into account in the analysis of this data.

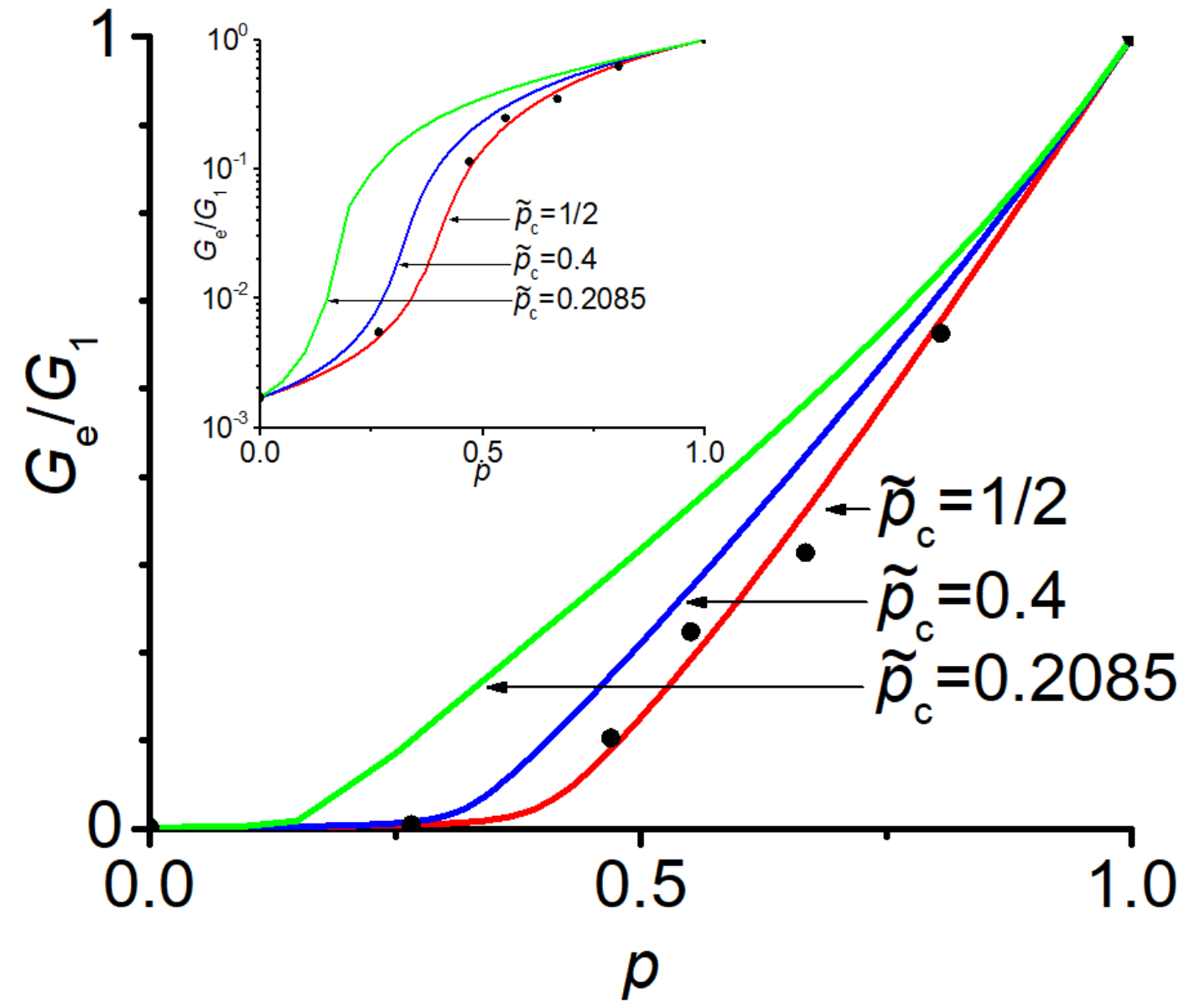

**Figure 7.** Normalized shear modulus of block copolymers of styrene and butadiene (filled circles). Experimental values are taken from [53]: $G_1/G_2 = 581.82$, $\nu_1 = 0.335$, $\nu_2 = 0.499$. The inset shows the same data on the logarithmic scale.

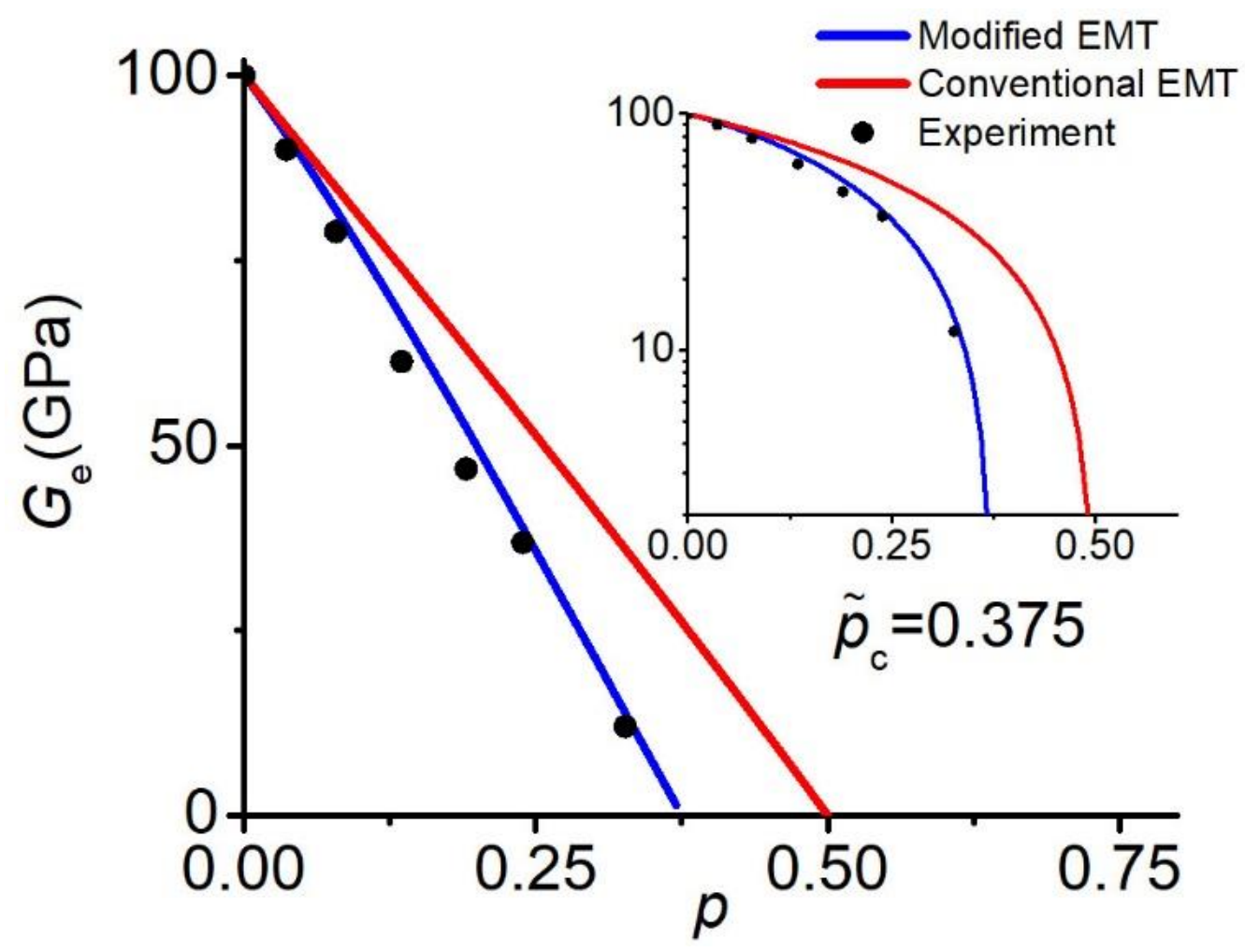

**Figure 8.** Porosity dependence of the effective shear modulus for porous $Th_2O$ (powder size 0 – 2 μm). Experimental values are taken from [55]. The inset shows the same data on the logarithmic scale.

In [55], porous materials were investigated ($G_1 = 0\,\mathrm{Pa}$). The percolation transition in the concentration dependence was conspicuous. It was found that the critical porosity (percolation threshold) depends on the powder size. The measured value of $p_c$ varied between 0.375 and 0.53 for $Th_2O$ powders. Figure 8 compares the results of both the conventional EMT and modified EMT with experimental results for a porous $Th_2O$ material (powder size 0 – 2 μm). A good agreement between the experimental results and the modified EMT, where $\tilde{p}_c = 0.375$, is observed. Moreover, it is clearly seen that the traditional EMT fails to describe this experiment. Similarly, our modified EMT is capable of describing the experimental elastic moduli for other powder sizes in [55].

To summarize, the theoretical results are supported by previous experiments, show improvement over the existing approach and may be used for validation of alternative approaches for determining the percolation threshold.

## 5. Conclusions

- A modification of the self-consistent EMT for the elastic properties of a random heterogeneous two-phase material is proposed. This modification allows one to prescribe any numerical value to a percolation threshold, which depends on the particular realization of a composite material. The modification does not change the values of the classical critical exponents ($f_{EMT} = S_{EMT} = 1$).
- A contradiction between the conductivity problem and the elasticity problem is eliminated. In a particular randomly inhomogeneous composite material, it is possible to simultaneously measure different physical properties, which should be determined by the same percolation threshold in an EMT.
- In the two-dimensional case, the proposed modification allows one to obtain an expression for the shear and Young's moduli at the percolation threshold, which symmetric with respect to the interchange of phases.

In the future work, the modification of the EMT for the elastic properties should allow us to employ the recently introduced concept of the moveable (field dependent) percolation threshold [49] to explain the giant change in the elastic moduli of the magnetorheological elastomers in moderate external magnetic fields, known as magnetorheolgical or field stiffening effect [56, 57]. These materials are also called magnetoactive elastomers when other physical properties are considered. The required dependence of the percolation threshold on the magnetic field was proposed empirically in [58].

**Author Contributions:** Conceptualization, A.S. and M.S.; methodology, A.S.; formal analysis, A.S.; investigation, P.Y., A.S and M.S.; writing—original draft preparation, A.S. and M.S.; writing—review and editing, M.S. and A.S.; visualization, P.Y.

**Funding:** The research of A.S. and P.Y. received no external funding. The research of M.S. was funded by the Deutsche Forschungsgemeinschaft (DFG, German ResearchFoundation), grant number 389008375.

**Acknowledgments:** The authors are grateful to Professor Viktor Kalita for discussion on the subject of this paper.

**Conflicts of Interest:** The authors declare no conflict of interest.

## Appendix A

### Calculation of the percolation threshold and the Poisson's coefficient in the traditional EMT for $G_1 \to \infty$.

First, the system of equations (7) is re-written as

$$\left.\begin{aligned}
&p\left[\frac{1}{1+\alpha_{\mathrm{e}}\left(\frac{G_1}{G_e}\cdot\frac{1+\nu_1}{1+\nu_e}\cdot\frac{1-2\nu_{\mathrm{e}}}{1-2\nu_1}-1\right)}-1\right]+(1-p)\left[\frac{1}{1+\alpha_{\mathrm{e}}\left(\frac{G_1}{G_e}\frac{1+\nu_2}{1+\nu_e}\frac{1-2\nu_{\mathrm{e}}}{1-2\nu_2}-1\right)}-1\right]=0\\
&p\left[\frac{1}{1+\beta_{\mathrm{e}}\left(\frac{G_1}{G_{\mathrm{e}}}-1\right)}-1\right]+(1-p)\left[\frac{1}{1+\beta_{\mathrm{e}}\left(\frac{G_2}{G_e}-1\right)}-1\right]=0
\end{aligned}\right\} \quad \text{(A1)}$$

Then, we introduce variables

$$m_{\mathrm{e}}=1/G_{\mathrm{e}}\,, m_1=1/G_1\,, m_2=1/G_2\,, \quad \text{(A2)}$$

and reformulate (A1) as

$$\left.\begin{aligned}
&p\left[\frac{1}{1+\alpha_{\mathrm{e}}\left(\frac{m_e}{m_1}\frac{1+\nu_1}{1+\nu_e}\frac{1-2\nu_{\mathrm{e}}}{1-2\nu_1}-1\right)}-1\right]+(1-p)\left[\frac{1}{1+\alpha_{\mathrm{e}}\left(\frac{m_e}{m_2}\frac{1+\nu_2}{1+\nu_e}\frac{1-2\nu_{\mathrm{e}}}{1-2\nu_2}-1\right)}-1\right]=0\\
&p\left[\frac{1}{1+\beta_{\mathrm{e}}\left(\frac{m_e}{m_1}-1\right)}-1\right]+(1-p)\left[\frac{1}{1+\beta_{\mathrm{e}}\left(\frac{m_e}{m_2}-1\right)}-1\right]=0
\end{aligned}\right\} \quad \text{(A3)}$$

At $m_1 \to 0$ the equations are written as

$$\left.\begin{aligned}
&\frac{1}{1+\alpha_{\mathrm{e}}\left(\frac{m_{\mathrm{e}}}{m_2}\frac{1+\nu_1}{1+\nu_{\mathrm{e}}}\frac{1-2\nu_{\mathrm{e}}}{1-2\nu_1}-1\right)}-1=\frac{p}{1-p}\\
&\frac{1}{1+\beta_{\mathrm{e}}\left(\frac{m_{\mathrm{e}}}{m_2}-1\right)}-1=\frac{p}{1-p}
\end{aligned}\right\}. \quad \text{(A4)}$$

Putting now $G_e \to \infty$ ( $m_e \to 0$) and $p \to p_c$, we find (compare with (9))

$$\alpha_{\mathrm{e}}=p_{\mathrm{c}},\quad \beta_{\mathrm{e}}=p_{\mathrm{c}} \quad \text{(A5)}$$

From where it immediately follows that

$$p_{\mathrm{c}}^{\mathrm{E}}=\frac{1}{2},\quad \nu_{\mathrm{e}}\left(p=p_{\mathrm{c}}^{\mathrm{E}}\right)=\frac{1}{5},\quad G_1=\infty. \quad \text{(A6)}$$

**Appendix B**

**Critical exponents in traditional and modified EMT.**

Using variables $m_{\mathrm{e}}=1/G_{\mathrm{e}}$ , $m_1=1/G_1$ , $m_2=1/G_2$ (A2), for $m_1 \to 0$ equations (7) can be written in the following form:

$$\left.\begin{array}{l} -\dfrac{3m_2-6m_2\nu_2-3m_2\nu_e+6m_2\nu_2\nu_e}{(2\nu_e-1)(2m_2+m_e-4m_2\nu_2+m_e\nu_e)}-1=\dfrac{p}{1-p} \\ \dfrac{15m_2-15m_2\nu_e}{7m_2+8m_e-5m_2\nu_e-10\nu_e m_e}-1=\dfrac{p}{1-p} \end{array}\right\}. \tag{B1}$$

From the first equation in the system of equations (B1) we obtain:

$$\nu_{\mathrm{e}}=\frac{m_2(-1+2\nu_2)+m_e(1+\nu_2)+3pm_2(1-2\nu_2)}{m_2(1-2\nu_2)+2m_e(1+\nu_2)+3pm_2(1-2\nu_2)}. \tag{B2}$$

Introducing a variable $\tilde{m}_e = m_e / m_2$ we get

$$\nu_e=\frac{\tilde{m}_e(1+\nu_2)+(3p-1)(1-2\nu_2)}{2\tilde{m}_e(1+\nu_2)+(3p+1)(1-2\nu_2)}. \tag{B3}$$

From the second equation of the system of equations (A2.1) we obtain:

$$\nu_{\mathrm{e}}=\frac{1}{5}\cdot\frac{8+7\tilde{m}_{\mathrm{e}}-15p}{2+\tilde{m}_{\mathrm{e}}-3p}. \tag{B4}$$

Equating the right-hand sides of solutions obtained

$$\frac{1}{5}\cdot\frac{8+7\tilde{m}_{\mathrm{e}}-15p}{2+\tilde{m}_{\mathrm{e}}-3p}=\frac{\tilde{m}_e(1+\nu_2)+(3p-1)(1-2\nu_2)}{2\tilde{m}_e(1+\nu_2)+(3p+1)(1-2\nu_2)} \tag{B5}$$

we get from a quadratic equation for $\tilde{m}_{\mathrm{e}}$

$$\tilde{m}_{\mathrm{e}}=\frac{14\nu_2-3p(1+3p\nu_2)-4\pm\sqrt{\xi}}{4(1+\nu_2)}, \tag{B6}$$

where

$$\xi=9p^2(3\nu_2+1)^2-12p(5\nu_2{}^2-7\nu_2+6)+4(5\nu_2-4)^2. \tag{B7}$$

Selecting the solution with the plus sign in the numerator of (B6) and using the method of Pade approximants for calculating the critical exponents, we find for $S$ in the EMT approximation

$$S_{\mathrm{EMT}}=\lim_{p\to 1/2^-}\left\{\left(p-\frac{1}{2}\right)\left[\frac{\partial}{\partial p}\ln\left(\frac{1}{m_{\mathrm{e}}}\right)\right]\right\}=1. \tag{B8}$$

Therefore, the critical exponent of the effective Young's and shear moduli below the percolation threshold in the framework of the EMT approximation is equal to unity: $S_{\mathrm{EMT}} = 1$. The numerically obtained value, which is outside of the framework of the EMT, is $S = 0.82$ [1,2,15,16].

The expression (B8) can be generalized for an arbitrary percolation threshold $\tilde{p}_{\mathrm{c}}$, where, obviously, the modification term (23) have to be taken into account in calculation of $m_{\mathrm{e}}$:

$$S=\lim_{p\to\tilde{p}_{\mathrm{c}}^-}\Sigma(p)=\lim_{p\to\tilde{p}_{\mathrm{c}}^-}\left\{\left(p-\tilde{p}_{\mathrm{c}}\right)\left[\frac{\partial}{\partial p}\ln\left(\frac{1}{m_{\mathrm{e}}}\right)\right]\right\}. \tag{B9}$$

The auxiliary function $\Sigma(p)$ can be easily calculated numerically for any concentration *p*, which is arbitrarily smaller than $\tilde{p}_c$. The Figure B1 (a) shows the exemplarily results of calculations. It is seen that $S$ = 1. Similar calculations can be performed for the critical exponent *f*:

$$f = \lim_{p\to\tilde{p}_c^+} \Phi(p) = \lim_{p\to\tilde{p}_c^+} \left\{ (p-\tilde{p}_c) \left[ \frac{\partial}{\partial p} \ln(G_e) \right] \right\}. \tag{B10}$$

The Figure B1 (b) shows the exemplarily results of calculations. It is observed that $f$ = 1.

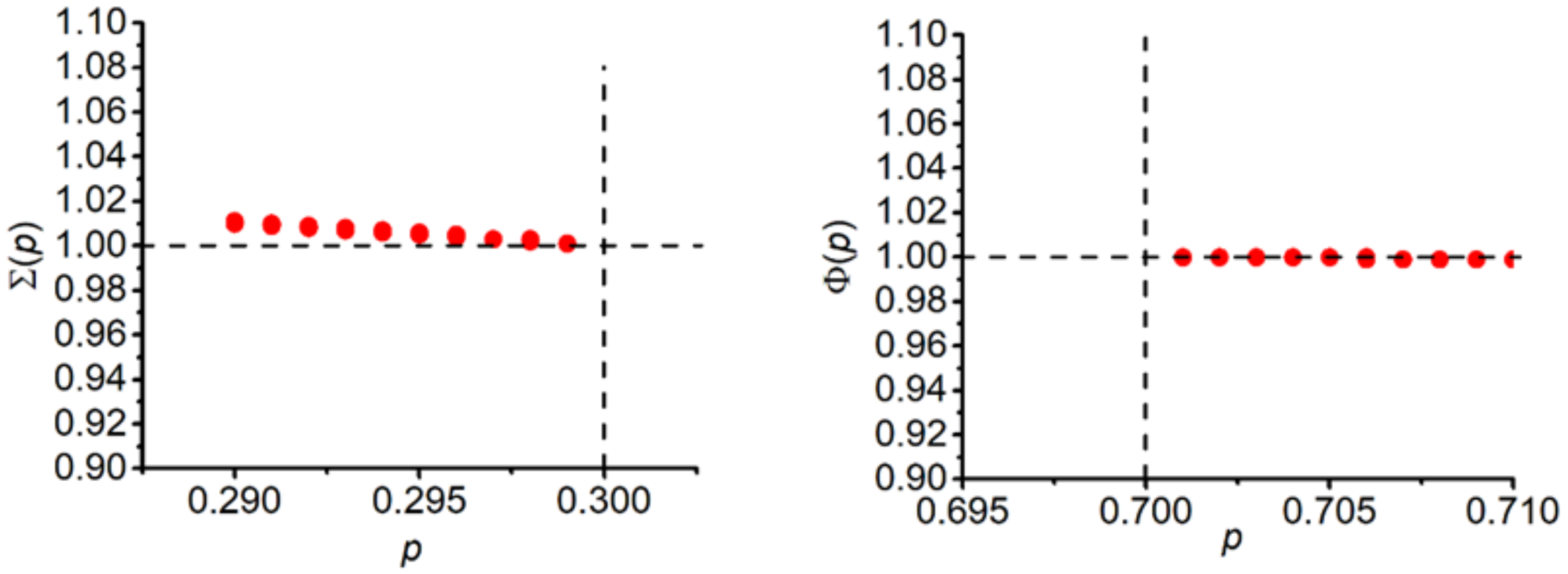

(**a**) (**b**)

**Figure B1.** (**a**) Concentration dependence of the function $\Sigma(p)$ for $p_c$ = 0.3 ($G_1$ = ∞); (**b**) Concentration dependence of the function $\Phi(p)$ for $p_c$ = 0.7 ($G_2$ = 0).

## Appendix C

**Calculation of the percolation threshold and the Poisson's ratio in the modified EMT with the SV term for $G_1 \to \infty$.**

Using variables $m_e = 1/G_e$, $m_1 = 1/G_1$, $m_2 = 1/G_2$ (A2) at $m_1 \to 0$ equations (22) can be written as:

$$\left. \begin{array}{l} \dfrac{1}{1+\alpha_e\left(\dfrac{m_e}{m_2}\dfrac{1+\nu_2}{1+\nu_e}\dfrac{1-2\nu_e}{1-2\nu_2}-1\right)} - 1 = \dfrac{p}{1-p-s(p,\tilde{p}_c)} \\ \dfrac{1}{1+\beta_e\left(\dfrac{m_e}{m_2}-1\right)} - 1 = \dfrac{p}{1-p-s(p,\tilde{p}_c)} \end{array} \right\}. \tag{C1}$$

Putting now $G_e \to \infty$ ( $m_e \to 0$) and $p \to p_c$, we find, similarly to Section 4.1.1, that

$$p_c = \tilde{p}_c, \quad \nu_e(p=\tilde{p}_c) = 1/5. \tag{C2}$$